\documentclass[aps, pre, twocolumn, showpacs, amsmath, amssymb, superscriptaddress, reprint, longbibliography]{revtex4-1}
\usepackage[utf8]{inputenc}
\usepackage[T1]{fontenc}
\usepackage{xcolor}
\usepackage{graphicx}

\definecolor{darkblue}{rgb}{0,0,0.6}
\definecolor{darkred}{rgb}{0.6,0,0}
\usepackage[colorlinks=true,urlcolor=darkblue, citecolor=darkblue, linkcolor=darkred]{hyperref}
\usepackage{siunitx}

\newcommand{\bv}[1]{{\boldsymbol #1}}

\begin{document}

\title{Colloidal transport in bacteria suspensions: from bacteria scattering to anomalous and enhanced diffusion}

\author{Antoine Lagarde}
\thanks{These authors contributed equally to this work}
\affiliation{Univ Lyon, ENS de Lyon, Univ Claude Bernard Lyon 1, CNRS, Laboratoire de Physique, F-69342 Lyon, France}

\author{Noémie Dagès}
\thanks{These authors contributed equally to this work}
\affiliation{Univ Lyon, ENS de Lyon, Univ Claude Bernard Lyon 1, CNRS, Laboratoire de Physique, F-69342 Lyon, France}

\author{Takahiro Nemoto}
\affiliation{Philippe Meyer Institute for Theoretical Physics, Physics Department, École Normale Supérieure \& PSL Research University, 24 rue Lhomond, F-75231 Paris Cedex 05, France}
\affiliation{Mathematical Modelling of Infectious Diseases Unit, Institut Pasteur, 25-28 Rue du Docteur Roux, 75015 Paris, France}

\author{Vincent Démery}
\affiliation{Gulliver, CNRS, ESPCI Paris, PSL Research University, 10 rue Vauquelin, Paris, France}
\affiliation{Univ Lyon, ENS de Lyon, Univ Claude Bernard Lyon 1, CNRS, Laboratoire de Physique, F-69342 Lyon, France}

\author{Denis Bartolo}
\affiliation{Univ Lyon, ENS de Lyon, Univ Claude Bernard Lyon 1, CNRS, Laboratoire de Physique, F-69342 Lyon, France}

\author{Thomas Gibaud}
\email{thomas.gibaud@ens-lyon.fr}
\affiliation{Univ Lyon, ENS de Lyon, Univ Claude Bernard Lyon 1, CNRS, Laboratoire de Physique, F-69342 Lyon, France}


%
%



\begin{abstract}
Colloids coupled to a bath of swimming cells generically display enhanced diffusion. This transport dynamics stems from a subtle interplay between the active and passive particles that still resists our understanding despite decades of intense research. Here, we tackle the root of the problem by providing a quantitative characterisation of the single scattering events between a colloid and a bacterium. Based on our experiments, we build a minimal model that quantitatively predicts the geometry of the scattering trajectories, and enhanced colloidal diffusion at  long times. This quantitative confrontation between theory and experiments elucidates the microscopic origin of enhanced transport. Collisions are solely ruled by stochastic contact interactions responsible both for genuine anomalous diffusion at short times and enhanced diffusion at long times with no ballistic regime at any scale.
\end{abstract}

\maketitle

\section{Introduction}

In nature, virtually all swimming microorganisms rely on interactions with particles dispersed in their natural environments. Prominent examples include  protists grazing on microscopic preys, and sperm cells fertilizing ovocites. In the labs, researchers have successfully put synthetic and living microswimmers to work to achieve a dynamics out of reach of equilibrium systems, including the assembly and actuation of  micromachines~\cite{Sokolov2010,Dileonardo2010,Maggi2016,Aubret2018},  topological-defect healing in colloidal crystals~\cite{Ramananarivo2019}, and  enhanced transport in non-Brownian suspensions. This latter line of research goes back to one of the earliest active--matter experiment~\cite{wu2000}. Investigating the diffusion of colloidal particles dispersed in a liquid film hosting swimming bacteria,  Wu and Libchaber laid out the foundation of active transport, and made seemingly simple observations that remain controversial despite twenty years of intense research~\cite{Bechinger2016}. 
 
When passive colloidal particles are dispersed in a dilute solution of motile organisms, they display a generic two-time dynamics. At long times,  regardless of the nature of the swimming particles, the multiple uncorrelated interactions between the active and passive units result in an enhanced diffusive dynamics characterized by a Gaussian displacement statistics~\cite{wu2000,Leptos2009, jeanneret2016, Ortlieb2019}. Micron-size colloids dispersed in a suspension of {\it E. coli} can diffuse as fast as nanoparticles in water. By contrast, at short times, the transport dynamics does not map to equilibrium and is generically  non-Gaussian and superdiffusive.  This anomalous dynamics is however non-universal and was the subject of contradictory reports. 
One situation was thoroughly investigated by the group of Polin, who established the ballistic nature of colloid transport in situations where the passive particles are much smaller than the active units~\cite{jeanneret2016,Mathijssen2018}. Far-field hydrodynamics play no role, and ballistic motion merely echoes  the persistent motion of the swimmers in carrying the colloidal particles.  Conversely, when the size of the passive objects  compares or exceeds that of the swimming particles, the situation remains  elusive and controvertial. Both the early experiments by Wu and Libchaber ({\it E. coli}), and the more recent results by Valeriani et al.\cite{valeriani2011} ({\it B. subtilis}) clearly demonstrated a non-ballistic regime at short time scales. However a plethora of theories and numerical simulations predict short-time ballistic transport with no consensus on the relative contribution of hydrodynamic and contact interactions, see e.g. ~\cite{Pushkin2014,Thiffeault2015,Morozov2014,Burkholder2017,Cugliandolo2015,pushkin2013} and references therein. To date, the only available explanation for the anomalous diffusion of passive particles coupled to  active baths relies on the emergence of collective motion and therefore does not apply to the vast majority of experiments performed in dilute suspensions~\cite{wu2000,Gregoire2001}. The current status is that most experiments on active transport are now analysed implicitly assuming a crossover between a ballistic and a diffusive regime~\cite{valeriani2011,Gasto2011,patteson2016}. The primary reason for this rather confusing situation is twofold. Firstly, we lack a clear characterization and understanding of the microscopic scattering dynamics ruling the couplings between active-swimmer baths and passive particles. Secondly, the low temporal resolution and small dynamical range of the control parameters  hinder the quantitative characterization of the asymptotic statistics in the current state-of-the-art experiments.
 
In this article, we rectify this situation investigating the enhanced transport of colloids in {\it E. coli} suspensions. We first provide a comprehensive characterization of the  collisions between a swimming bacteria and passive colloidal beads, and introduce a minimal theoretical model that faithfully account for their full scattering dynamics. Our model rules out the impact of far-field hydrodynamic interactions. Investigating the consequences of this scattering process on active transport, we firmly establish the existence of genuine superdiffusion and non Gaussian transport  at short time,  revealing a complex interplay between the propulsion of the active units and the displacements of the passive colloids upon physical collisions. 

\section{Experimental setup}
\label{sec:bact}
\begin{figure}
	\centering
    \includegraphics[width=0.8\linewidth]{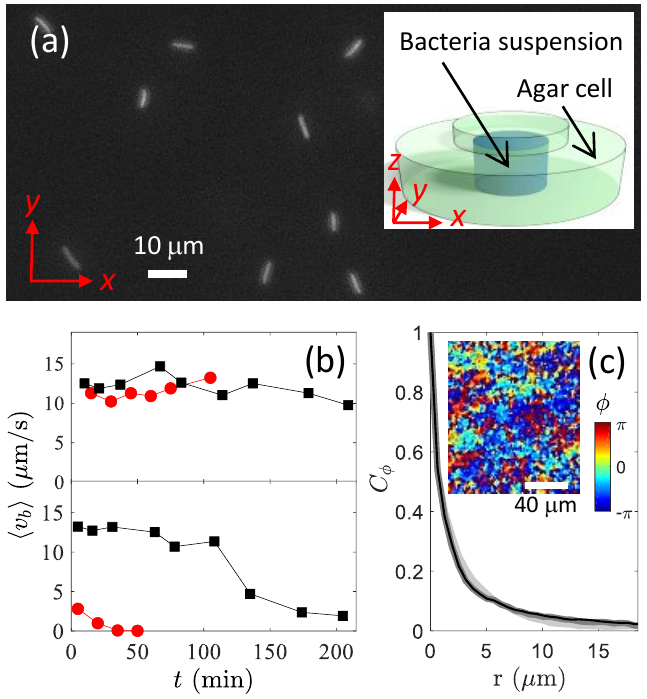}
      \caption{Bacteria suspension. (a) Fluorescence microscopy image of a dilute bacteria suspension. Inset: Sketch of the agar observation cell. (b) Average bacteria velocity $v_{\rm b}$ as a function of the age of the solution  in the agar cell (square) and in a standard cover-slip observation cell (circle) for two bacteria concentrations $c=1.8$ $10^{10}$mL$^{-1}$ (top) and $c= 18$ $10^{10}$mL$^{-1}$ (bottom). (c) Correlation function of  the velocity-field orientation $\phi$ plotted as a function of the distance $r$ for concentrations $c=2$ (light grey), 10 (dark grey) and 20$~10^{10}$mL$^{-1}$ (black).  
Inset: The instantaneous orientation shows very little spatial correlation ($c= 18$ $10^{10}$mL$^{-1}$).
}
    \label{f:exp}
\end{figure}

We provide a thorough description of our experiments in the Methods section~\ref{sm}. In brief, our experimental system is composed of an aqueous dispersion of bacteria seeded with polystyrene colloids of radius $\ell_{\rm c}=5~\mu\rm m$ (if not specified otherwise). The Bacteria are fluorescent smooth-runner mutants RP437 of {\it E. coli}. They are smaller that the passive colloids, measurements of their  average diameter is $0.5\,\rm \mu m$ and their average length is $L_{\rm b}=4.3\pm2\,\rm \mu m$, values comparable to the literature~\cite{schwarz2016}.  Bacteria suspensions are prepared according to the  protocols reviewed in~\cite{schwarz2016}. Using standard single particle tracking~\cite{Crocker1996}, we find that the bacteria  swim at an average speed  $ v_{\rm b}=15 \pm 4$ $\mu$m/s. In order to run experiments  long enough  to achive large-enough  statistics, the bacteria solutions are placed in  agar observation cells sketched in Fig. \ref{f:exp}(a). The primary advantage of agar-based devices is that the bacteria remain alive and active with stationary dynamics over more than 100 min, in contrast to standard glass or PDMS cells where the  bacteria average velocity decays monotonically and can vanish in  $\sim 10\,\rm min$, see Fig. \ref{f:exp}(b). Using particle image velocimetry (PIV) on bright field images, we measure the velocity field in pure bacteria suspensions. Varying the bacteria  concentrations $c$ from $10^7$ to $3\times10^{11}$ mL$^{-1}$, we never observe spatial correlations in the active flows, Fig. \ref{f:exp}(c): the bacteria dispersion is an isotropic active fluid at all concentrations considered in this work. 

\section{Bacteria-colloid collisions}

\subsection{Experimental results}
\label{sec:collision}
\begin{figure}
	\centering
    \includegraphics[width=7.cm]{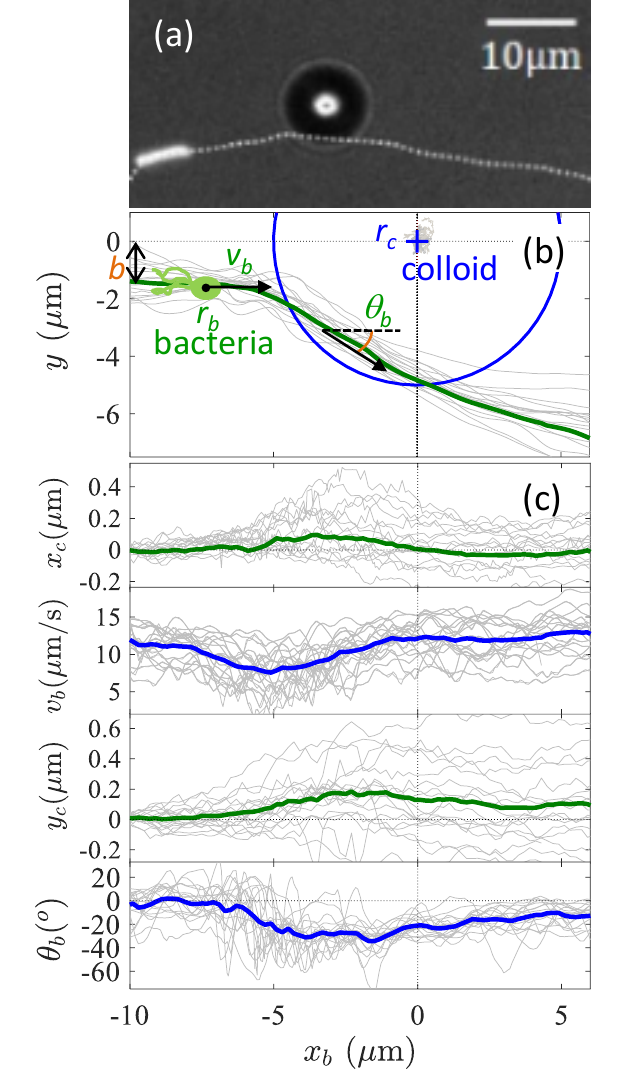}
      \caption{Collision between a bacterium and a colloid. (a) Typical microscopy image of a collision. (b) The bacteria at position ${\bv r}_{\rm b}(x_{\rm b}, y_{\rm b})$ has a velocity $v_{\rm b}$. $b$ is the $y-$distance to the center of the bacteria taken at $x_{\rm b}=-10~\mu$m. At $x_{\rm b}=-10~\mu$m the velocity $v_{\rm b}$ is oriented along the $x-$axis. The colloid  of radius $\ell_{\rm c}= 5~\mu$m is characterized by the position of its center of mass ${\bv r}_{\rm c}(x_{\rm c}, y_{\rm c})$. The bacteria trajectory is deviated by the colloid by an angle $\theta_{\rm b}$. Grey lines indicate individual quantities and the green or blue lines are the average quantities. (c) Collision parameters $x_{\rm c}$, $y_{\rm c}$, $v_{\rm b}$ and $\theta_{\rm b}$ as a function of $x_{\rm b}$. 
}
    \label{f:choc1}
\end{figure}

We start by analysing the collisions between a single colloid of radius $a=5\mu$m and a single bacterium, see Fig.~\ref{f:choc1}(a). 
Combining bright field  and fluorescence microscopy makes it possible to simultaneously track  instantaneous positions of the colloid, ${\bv r}_{\rm c}(t)=(x_{\rm c}(t),y_{\rm c}(t))$, and of the colliding bacterium, ${\bv r}_{\rm b}(t)=(x_{\rm b}(t),y_{\rm b}(t))$. 
To characterize the collision process, we choose the time origin when the bacterium and the colloid are separated by a distance of 10 $\mu$m. The spatial origin and orientation of the frame are then set so that ${\bv r}_{\rm c}(t=0)=0$, and $\hat{\bv y}\cdot {\bv v}_{\rm b}(t=0)=0$, where ${\bv v}_{\rm b}(t)$ is the instantaneous bacterium velocity,  see Fig.~\ref{f:choc1}(b). The impact parameter of the collision is then defined as $b=y_{\rm b}(t=0)$, Fig.~\ref{f:choc1}(b). 

Taking the bacterium position along the $x$-axis, $x_b$, as the parameter for both the bacterium and colloid trajectories, we monitor the displacement of the colloid ($x_c$, $y_c$) as well as the speed $v_b$ and deviation angle $\theta_b$, Figs.~\ref{f:choc1}(b) and~\ref{f:choc1}(c). 
Comparing the scattering trajectories of hundreds of bacteria, and grouping those corresponding to the same impact parameter $b$, we identify a set of robust features.  Both the colloids and the bacteria trajectories fluctuate around well-defined average path.  For a parameter $b=-1.5$~$\mu$m, the average path of both the bacterium and the colloid is clearly affected by the collision. The bacterium tends to push the colloid: $x_c>0$ and $y_c>0$ and the colloid slows down the bacterium and deviate its trajectory by an angle $\theta_b<0$ (when $b<0$). 

\begin{figure}
	\centering
    \includegraphics[width=7.5cm]{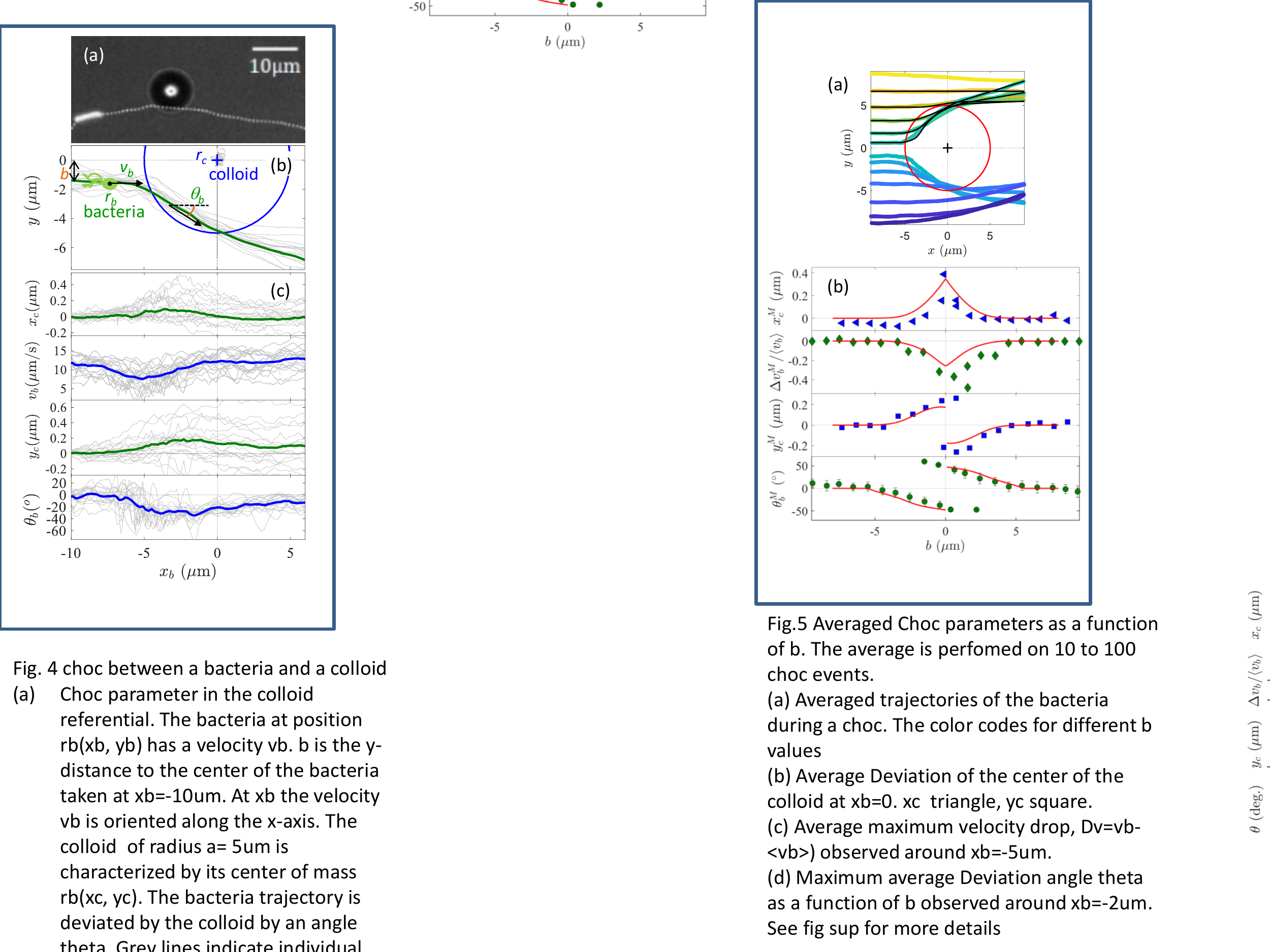}
      \caption{Averaged collisions as a function of the impact parameter $b$. The average is performed on 10 to 100 collision events. (a) Averaged trajectories of the bacteria during a collision. The color codes for different $b$ values.
      (b) Maximum deviation of the center of the colloid ($x^{\rm M}_{\rm c}$  (triangle), $y^{\rm M}_{\rm c}$ (square)),
      maximum relative velocity drop, $\Delta v_{\rm b}/\langle v_{\rm b} \rangle$ (diamond)
      and maximum deviation angle $\theta_{\rm b}^{\rm M}$ (circle) as a function of $b$; the maxima are computed on the averaged trajectories.
      Red lines are model predictions.
      }
    \label{f:choc2}
\end{figure}

Given the radial symmetry of the colloidal particles, the average dynamics is accurately determined by the sole impact parameter $b$, Fig.~\ref{f:choc2}(a). In order to quantify this scattering, we plot, as a function of $b$, in Fig.~\ref{f:choc2}(b), the maximum displacements of the colloid along the $x$ and $y $ directions ($x_{\rm c}^{\rm M}$, and $x_{\rm c}^{\rm M}$) as well as the maximum velocity drop $\Delta v_b^{\rm M}=v_b^{\rm M}-\langle v_b \rangle$ and deviation angle $\theta_b^{\rm M}$ of the bacterium trajectories. Those maxima are reached around $x_b\sim -2~\mu$m, except for the velocity drop which takes place around $x_b\sim -5~\mu$m, Fig.~\ref{f:choc1}(c).

From Fig.~\ref{f:choc2}(b), we can readily infer four essential results. (\textit{i}) We find no average displacement of the colloidal particle when the impact parameter exceeds one colloid radius: the interactions between the colloids and the bacteria are short ranged. We can therefore discard the role of far-field hydrodynamic interactions in the collision process.  (\textit{ii}) On the contrary, for $|b|<5$~$\mu$m, the bacterium and colloid trajectories are affected. The scattering of each bacterium is systematically associated to a net displacement of the colloidal bead over distances significantly larger than its typical diffusion length over the collision time. The colliding bacterium pushes the colloid away from its initial position. We indeed find that  $x_{\rm c}^{\rm M}$ is positive and  that $y_{\rm c}^{\rm M}$ and $b$ have opposite signs for all impact conditions. The effect is maximal for a frontal collision, $b\sim 0$. (\textit{iii}) The bacterium slows down upon contact and the reduction of its swimming speed $\Delta v_b^{\rm M}$ mirrors the magnitude of the colloid displacements. This effect is also maximal for $b\sim 0$, where the relative velocity drop is around $40 \%$. (\textit{iv}) The bacterium is mostly scattered upward if $b>0$ and downward if $b<0$. The scattering angle of the bacterium trajectories $\theta_b^{\rm M}$ is maximal upon head-on collisions ($b\sim0$) and reaches a value of $50^\circ$. We never observe any orbital trajectory akin to that observed using synthetic active colloids or when {\it E. coli} collides cylindrical posts~\cite{takagi2014,spagnolie2015,Sipos2015}. These last two observations further confirm the prominence of contact or lubrication interactions in the collision dynamics. This vouch for steric repulsion between the colloid and the bacterium and exclude hydrodynamic interactions, which would leads to an effective attraction along the $y$-direction due to the pusher nature of the swimmer~\cite{Shum2017}.

\subsection{Theoretical description of bacteria--colloid scattering}
\label{sec:model}

\begin{figure*}
	\centering
    \includegraphics[width=17.5cm]{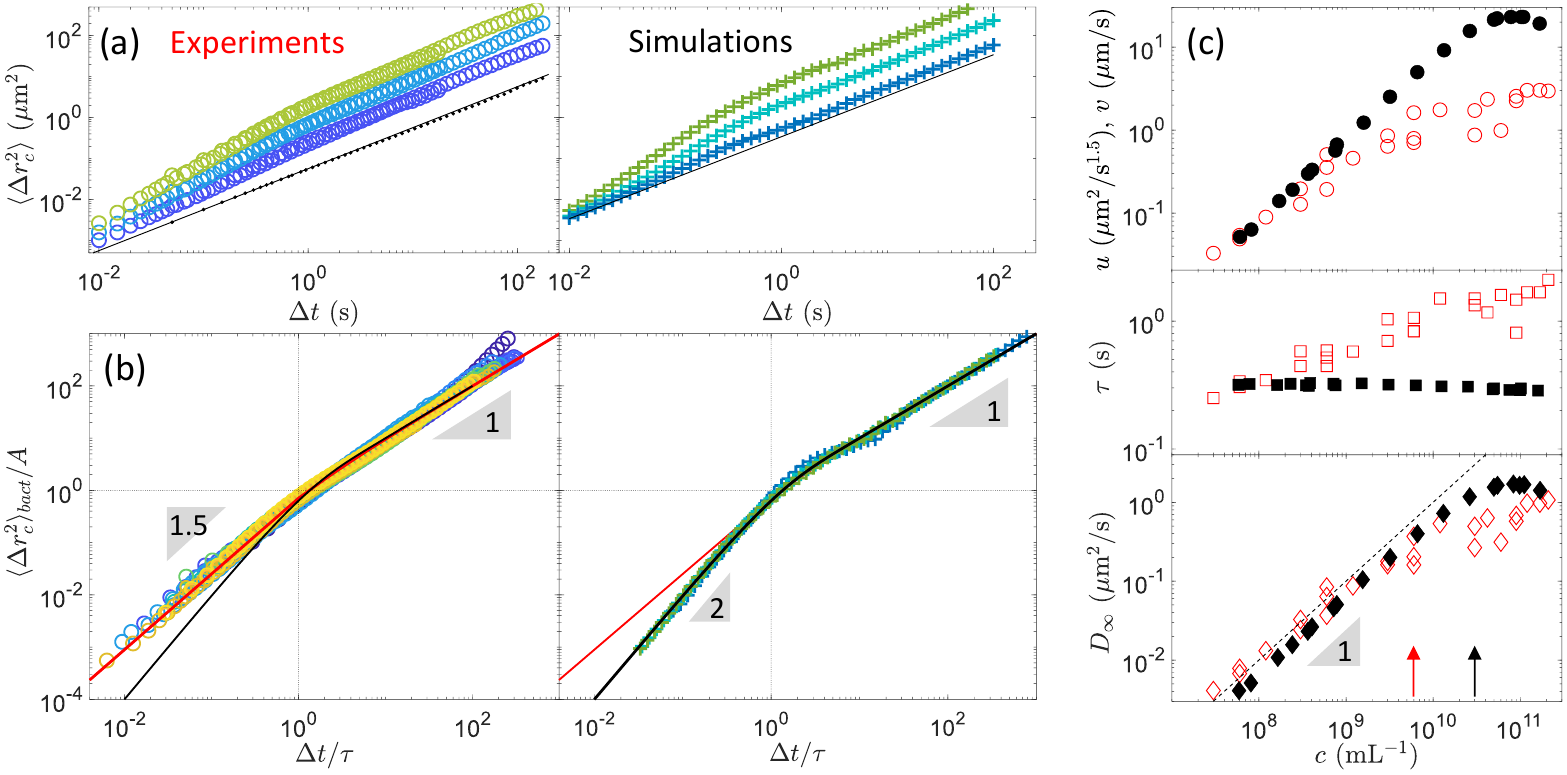}
      \caption{Dynamics of colloids in a bath of bacteria, experiments versus simulation. (a) Mean square displacements of the colloids in a bath of bacteria at concentrations $c=0.6$, 6 and  $90 \cdot 10^9$mL$^{-1}$ (experiments) and $c=0.7$, 6 and  $50 \cdot 10^9$mL$^{-1}$ (simulations). For the simulations:  $\ell_{\rm b}=0.5\mu \rm m$ (for bacteria) and $\ell_{\rm c}=5\mu \rm m$ (for colloids).      The black line corresponds to the free diffusing colloids. (b) $\langle \Delta r_{\rm c}^2 \rangle_\text{bact}=\langle \Delta r_{\rm c}^2 \rangle -4D_0 \Delta t$ rescaled  for different concentrations of bacteria  ranging from $c=3\cdot 10^7$ to $2\cdot 10^{11}$ mL$^{-1}$. The red line is an empirical fit with $\langle \Delta r_{\rm c}^2 \rangle_\text{bact}=u \Delta t^{1.5}/(1+\Delta t/\tau)^{0.5}$ with $D_{\infty}=u/\tau$ and $A=u\tau^{1.5}$. The black line is the fit proposed by Wu et al. \cite{wu2000}: $\langle \Delta r_{\rm c}^2 \rangle_\text{bact}=4D_{\infty} \Delta t(1-e^{-\Delta t/\tau})$ with $v=4D_{\infty}/\tau$ and $A=4D_{\infty}$. The color codes for increasing  bacteria concentration $c$ from blue to yellow. (c) Scaling of the fit parameters $u$, $v$, $\tau$ and $D_{\infty}$ as a function $c$: red (experiments) and black (simulations). Arrows indicate the concentration $c^*$ that sets the upper limit of the linear regime for $D_{\infty}$.
}
    \label{f:msd}
\end{figure*}

In order to account for our planar optical measurements,  we describe the 3D collision between a bacterium and a colloid by an effective two-dimensional model 
using the collision dynamics of two disks with different radii,  $\ell_{\rm b}=0.25~\mu \rm m$ (for the bacterium) and $\ell_{\rm c}=5~\mu \rm m$ (for the colloid), positioned at $\bv r_{b}$, $\bv r_{c}$.
Their dynamics are described by the two  over-damped equations:
\begin{equation}
\frac{d \bv r_{b}}{dt} = v_{b}\hat {\bv e} +  \frac{1}{\gamma_{\rm b}} \bv F_{\rm int}^{b} , \quad
\frac{d \bv r_{c}}{dt} =  \frac{1}{\gamma_{\rm c}} \bv F_{\rm int}^{c}.
\label{eq:1}
\end{equation}
Here $\gamma_{\rm c}=6 \pi \mu \ell_{\rm c}$ is the drag constant of a spherical colloid given by Stoke's law, where $\mu=10^{-3}~\si{\pascal\second}$ is the dynamic viscosity of water, and $\gamma_{\rm c}\simeq 2.9\cdot 10^{-8}~\si{N s / m}$ is the bacterium longitudinal drag coefficient\cite{chattopadhyay2006}.
We assume an isotropic drag coefficient for the bacterium to keep the model as simple as possible.
In Eq.~\eqref{eq:1}, $v_{b}=15\,\mu {\rm m}/s$ is the typical velocity of the bacteria far from the colloid and $\hat {\bv e} = (\cos \theta,\sin \theta)$ is a unit vector that defines its orientation. We model the bacterium-colloid repulsive interaction by the  forces $\bv F_{\rm int}^{b},\bv F_{\rm int}^{c}$ both deriving from the same contact potential
\begin{equation}
\bv F_{\rm int}^{c,b} = -\frac{\partial }{\partial \bv r_{c,b}} U(|\bv r_{c}-\bv r_{b}|),
\end{equation}
where $U(r)$ is the WCA potential of range $r_0 = 2^{1/6}(\ell_{\rm b} + \ell_{\rm c})$ and magnitude $\epsilon$:
\begin{equation}
U_{\rm WCA}(r) =  \epsilon \left[ \left(\frac{r_0}{r} \right)^{12}-2\left(\frac{r_0}{r} \right)^{6} +1\right]\Theta(r_0-r),
\end{equation}
Since we aim at understanding the average properties of the collision, we do not introduce noise in our model, while noise is obviously present in the experiments as can be seen from the dispersion of the (grey) trajectories in Fig.~\ref{f:choc1}.
Guided by our  experimental observations, we also introduce a torque that aligns the orientation of the bacterium $\hat{\bv e}$ with the surface of the colloid during a collision, Figs.~\ref{f:choc1}(b) and~\ref{f:choc2}(b).
Denoting by $\theta_{\rm tan}$ the angle between $\hat {\bv e}$ and the vector tangent to the colloid surface, the overdamped orientational dynamics of the bacterium reads
\begin{equation}
\frac{d\theta}{dt} = \tau  \sin(\theta_{\rm tan})  \Theta(r_0 - |\bv r_{c}-\bv r_{b}| ),
\label{eq:orientation}
\end{equation}
where  $\tau$ is the torque magnitude,  and $r_0$ is the effective range of the aligning torque, taken identical to that of $U({\mathbf r})$ for the sake of simplicity. We note that Eq.~\ref{eq:orientation} differs from the models introduced in~\cite{spagnolie2015,Sipos2015} where the angular dynamics selects a preferred finite angle with the solid surface, thereby promoting circular orbits. As we observe no orbiting trajectory we neglect this contribution.

We determine the parameters $\epsilon=0.071\,k_{\rm b}T$ and $\tau=\SI{3.35}~{s^{-1}}$ by fitting the model to the experimental data of single collision trajectories as a function of the impact parameter $b$, Fig.~\ref{f:choc2}(a). 
Using these fitting parameters, we calculate the collision parameters $x_{\rm c}$, $y_{\rm c}$, $\Delta v_{\rm b}/v_{\rm b}$, $\theta_{\rm b} $ for all values of $b$. 
Note that the small repulsion parameter allows for the bacterium to ``penetrate'' into the colloid (Fig.~\ref{f:choc2}(a)); this comes from the projection of three dimensional trajectories onto the observation plane, whereby bacteria passing above or below the colloid are pictured ``inside'' the colloid.
As shown in Fig.~\ref{f:choc2}(b), the trajectories predicted from this minimal model quantitatively captures  the average properties of the collision for all impact parameter values. We note that the model remains robust upon small variations of the bacteria radii.

\section{Colloids dynamics as a function of the bacteria concentration}
\label{sec:msd}

\subsection{Mean square displacement}\label{}

We now turn our attention to the enhanced transport  dynamics of the passive colloids animated thermal fluctuations and collisions with the swimming cells. The colloid concentration is kept very low to avoid colloid--colloid interactions. To characterise the colloids motion, we record movies of colloids dispersed in bath of bacteria at a concentration $c$ and we track their trajectory ${\bv r}_{\rm c}$ as a function of time. At $c=0$, the colloids are weakly Brownian: the mean square displacement (MSD) of the colloids evolves linearly with the lag time $\Delta t$ as shown in Fig.~\ref{f:msd}(a), and we measure a free diffusion coefficient of $D_0=0.015~\mu$m$^2$/s, lower than the bulk diffusion coefficient as the colloids are sedimented on  the bottom surface of our observation chamber.

\begin{figure}
	\centering
    \includegraphics[width=8cm]{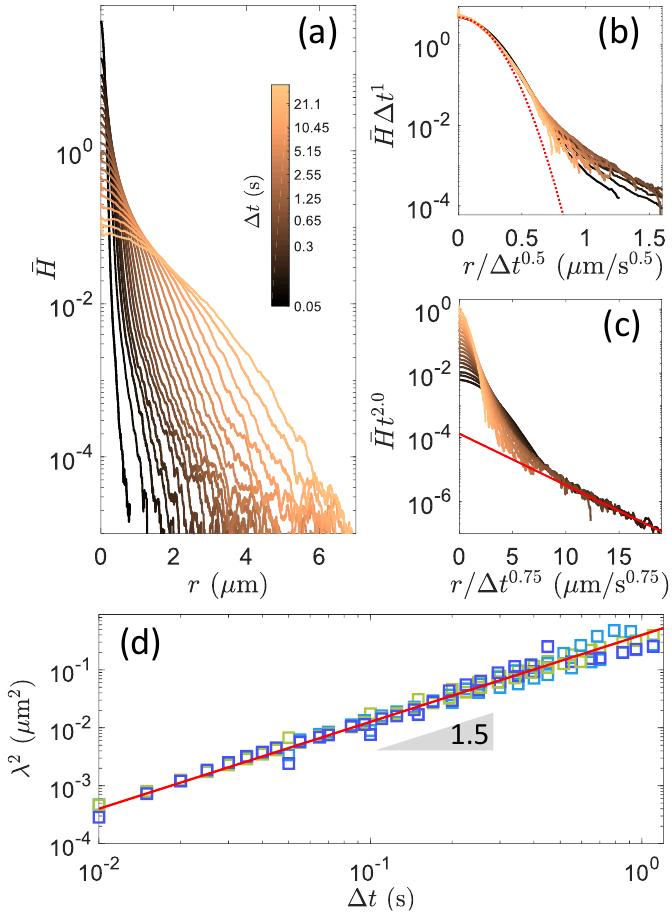}
      \caption{Scaling of the probability distribution function of a colloid in a bath of bacteria at concentration $c=1.2\cdot 10^8$ mL$^{-1}$. (a) Angular average of the self part of the Van Hove function $\bar H=PDF_{|\Delta r_{\rm c}|}/(2\pi r)$ at different lag times $\Delta t$. (b) $\bar H$ where $r$ is rescaled by $\Delta t^{0.5}$. The red line corresponds to $\bar H$ for colloid free diffusion ($c=0$). (c) $\bar H $ where $r$ is rescaled by $\Delta t^{0.75}$. The red line is an exponential fit. (d) Square of the exponential characteristic length $\lambda$ as a function of the time increment $\Delta t$  for 3 concentrations of bacteria $c=$0.6, 6 and  90.10$^9$mL$^{-1}$. The red line is a power law of exponent 1.5.
}
    \label{f:pdf}
\end{figure}

As shown in~Fig.~\ref{f:msd}(a), upon increasing the bacteria concentration, the MSD becomes larger as $c$ increases. In order to pinpoint the effect of the bacteria, we henceforth subtract the thermal contribution to the MSD and plot $\langle \Delta r_{\rm c}^2 \rangle_\text{bact}=\langle \Delta r_{\rm c}^2 \rangle -4D_0 \Delta t$.
Remarkably, a rescaling of the displacements by a constant $A(c)$ and lag time by a time scale $\tau(c)$ collapses all curves on a single master curve,  in Fig.~\ref{f:msd}(b).  This master curve  distinguishes two distinct asymptotic dynamics and can be empirically fitted by the function $\langle \Delta r_{\rm c}^2 \rangle_\text{bact} = u \Delta t^{1.5}/(1+\Delta t/\tau)^{0.5}$ with $D_{\infty}=u \tau^{0.5}/4$. This function interpolates between a long time diffusive dynamics where $\langle \Delta r_{\rm c}^2 \rangle_\text{bact}=4D_{\infty}\Delta t$ and a short time superdiffusive behavior  where $\langle \Delta r_{\rm c}^2 \rangle_\text{bact}=u \Delta t^{1.5}$ consistent with the early observations of Wu and Libchaber. Superdiffusion is observed over two orders of magnitude where all our experimental data collapse on the same master curve. The consistency of our short time observations  dismiss the hypothesis where this regime would be a mere  crossover from a ballistic to a diffusive dynamics \cite{wu2000}. In addition, we stress that the master curve accurately describes the transport dynamics  of colloids  at low bacteria concentration which further  dismiss the hypothesis that the anomalous scaling law $\langle r_{\rm c}^2 \rangle \sim \Delta t^{1.5}$ could be the fingerprint of collective motion~\cite{Gregoire2001}.

Fig.~\ref{f:msd}(c) shows the fitting parameters $u$, $\tau$ and $D_{\infty}$ as a function of the bacteria concentration $c$. Those parameters increase with $c$ and tend to plateau above $c^*\sim 6 \cdot 10^9$ mL$^{-1}$ which corresponds to a bacteria volume fraction of $\sim 2$\%. Only the value of $D_{\infty}$ was reported in the literature and our measurements are  in agreement with Refs.~\cite{jepson2013, pushkin2013}.
The linear evolution of $D_{\infty}$ with $c$ indicates that the diffusive process is additive: all collisions contribute independently to the dynamics. 
For $c>c^*$, however, a number of bacteria collide the colloid at once, the diffusive process is no longer additive and $D_\infty$ saturates to a finite value.



To gain more insight on  the active transport, we simulate the dynamics of a single colloid in a bacteria bath using the model constructed in section~\ref{sec:model}, and neglecting the interactions between bacteria.
Noticeable similarities and differences  with our experiments illuminate the nature of the origin of enhanced transport, ~Fig.~\ref{f:msd}.
As in experiments, the MSD can be collapsed on a single master curve $\langle \Delta r_{\rm c}^2 \rangle_\text{bact}=4 D_{\infty} \Delta t(1-e^{-\Delta t/\tau})$~\cite{wu2000}.
The fitting parameters $D_{\infty}$, $\tau$ and $v=4D_{\infty}/\tau$ follow the exact same trend as in our expeirments:  they increase with $c$ and tend to plateau above $c^*\sim 2 \cdot 10^{10}$~mL$^{-1}$ which corresponds to a volume fraction of $\sim 6.8$\%.
In the linear regime, experiment and simulations  quantitatively agree on $D_{\infty}$, indicating that the long-time enhanced diffusion is fully captured by the average scattering dynamics of the bacteria.

The two main discrepancies between the simulations and the experiments are even more insightful. Firstly,  the concentration $c^*$ where the diffusion coefficient departs from the linear regime is much higher for simulations ($c^*\sim 2~10^{10}$~mL$^{-1}$) and $D_\infty$ is larger in simulations at high $c$. These observations indicate that the bacteria-bacteria interactions absent in the simulations are chiefly responsible for the saturation of the effective diffusivity observed in our experiments. Secondly, the short time dynamics are qualitatively different. The MSD varies ballistically in the simulations (Fig. \ref{f:msd}(b)), at odds with the anomalous scaling found in the experiments. This essential difference points towards the crucial role played by the fluctuations in the bacterial dynamics on short-time superdiffusion.


\subsection{Probability density function of the displacements}\label{}

\begin{figure}
	\centering
    \includegraphics[width=7.cm]{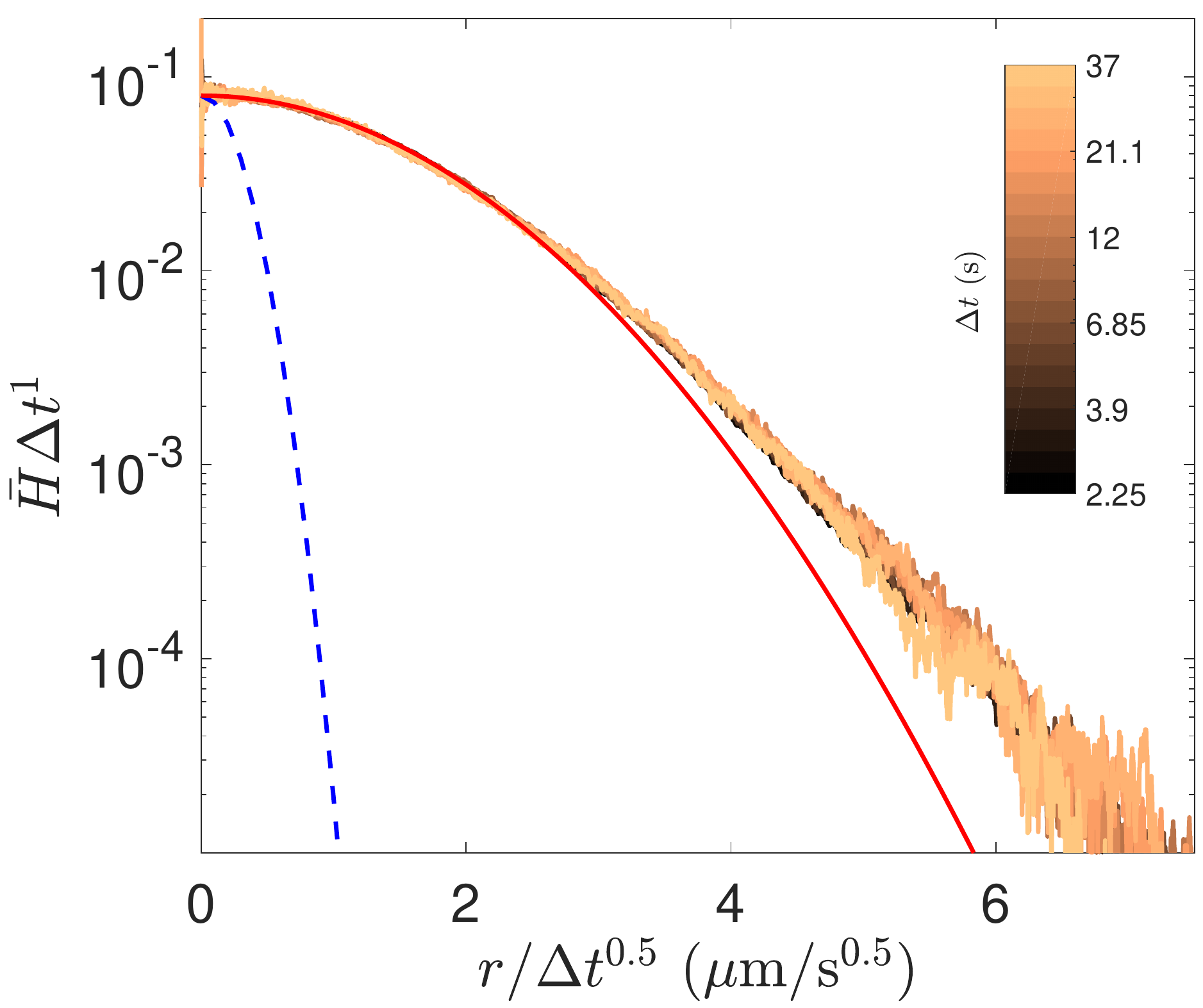}
      \caption{Scaling of $\bar H$ as a function of $r/\Delta t^{0.5}$ for $\Delta t>\tau$ in a bath of bacteria at $c= 90$ 10$^9$mL$^{-1}$. Dash blue line: free diffusion of the colloids ($c=0$). Red line: best Gaussian fit for $r/\Delta t^{0.5}< 2$~$\mu$m/s$^{0.5}$.}
\label{fig:s8}
\end{figure}

To further elucidate the anomalous dynamics of the colloids, we analyse the probability density function (PDF) of the  the colloid displacements. More quantitatively, we compute the angular average of the self part of the Van Hove function $\bar H=PDF_{|\Delta r_{\rm c}|}/(2\pi r)$ at different lag times $\Delta t$ (Fig.~\ref{f:pdf}), see section \ref{sm} for a detailed definition.
$\bar H$ features a Gaussian center and exponential tails, which are more prominent at short times for all concentrations.
At short times and  low concentrations, collisions are rare events, and the Gaussian center can be unambiguously attributed to thermal noise; indeed its standard deviation $\sigma$ is given by $\sigma^2\simeq 4D_0\Delta t$ (Fig.~\ref{f:pdf}(b)).
At long times, many collisions have occurred and the entire distribution becomes Gaussian as a consequence of the central limit theorem; the width of the distribution is $\sigma^2\simeq 4D_\infty \Delta t$  (Fig.~\ref{fig:s8}).

The exponential tails are characterized by their characteristic length $\lambda$.
For all concentration, at short times this length scales as $\lambda^2\sim \Delta t^{1.5}$ (Fig. \ref{f:pdf}(c)), corresponding to the anomalous scaling of the MSD. This behavior is yet another confirmation of the existence of a genuine anomalous transport regime at short times.
The numerical prefactor does not depend on the concentration (Fig. \ref{f:pdf}(d)), indicating that the exponential tails are due to the stochasticity of single-collision events, in agreement with the fact that  colloids experience an average number of collisions smaller than 1 for lag times smaller than $\tau\sim 1~\si{s}$. This essential observation further demonstrates that the anomalous active transport of passive colloids dispersed in a bath of swimmers is determined by the stochastic dynamics of the swimmer-colloid interactions and therefore cannot be captured by any deterministic interaction model.

\section{Concluding remarks}

The pioneering discussion of active transport by Wu and Libchaber\cite{wu2000} led to a simple picture akin to conventional Brownian motion\cite{Huang2011, Hammond2017}.
Bacteria have long been thought as playing the role of a heat bath leading to long time diffusion, and short time ballistic motion.
Investigating the scattering of a single bacterium with isolated colloids, we establish that the persistence of bacteria motion does not translate in a mere ballistic displacement of the passive particles.
In stark contrast, the subtle interplay between the propulsion of the swimming cell and the colloid displacement yields a genuinely anomalous and non-Gaussian dynamics.
In addition, combining experiments and theory we elucidate the origin of enhanced transport showing that it chiefly relies on contact interactions with imperceptible far-field hydrodynamic contributions.

The qualitative difference in the transport of colloidal bodies when  activated by collisions with different swimming cells is the most prominent when comparing our experiments to that of \cite{Leptos2009, jeanneret2016} where algae literally carry colloids with their swimming appendages.
This diversity of microscopic interactions translates in fundamental differences in the transport statistics thereby suggesting a wealth of design strategies for cell-powered microscopic motors and heat engines\cite{Krishnamurthy2016}.

\section{Methods}
\label{sm}

\indent \textbf{Microscopy} -- Microscopy measurements where performed with an inverted microscope (Ti-eclipse from Nikon). Images where recorded with CMOS camera (ORCA-Flash 3.0 from Hamamatsu). To visualize simultaneously the bacteria and the colloids we use fluorescence and bright field. In bright field, colloids act as a lens and focus the light in their center which enable us to track them. The light intensity of the bright field is tuned so that bacteria can also be seen simultaneously in fluorescence with gfp compatible filters.
\vspace{.1 cm}

\textbf{{\it E. coli} RP437} -- We use a mutant Escherichia Coli bacteria. The strain we use is {\it E. coli} RP437. The bacteria is modified to produce  GFP (Green Fluorescent Protein) so that it is fluorescent. The bacteria is also modified to become a smooth runner, with a high persistence length, i.e. with a long ballistic movement and rare tumbling episodes.
\vspace{.1 cm}

\textbf{{\it E. coli} preparation} -- {\it E. coli} are stored in a -80$^{\circ}$C freezer, in water (33 \% weight) and glycerol (66 \% weight). First we place a small amount of this initial mixture on an sterile agar plate (1.5 \%w of Aagar, 1 \%w of NaCl, 1 \%w of Tryptone, 0.5  \%w of Yeast Extract), with ampicillin, an antibiotic that allows us to select only our mutant. Then we put this plate in an incubator at 37$^{\circ}$C over the night, during which colonies originating from a single bacterium are formed. Then, an isolated colony is taken and dispersed in a liquid growth medium (1 \%w of NaCl, 1 \%w of Tryptone, 0.5  \%w of Yeast Extract in deionized water), in a tube permeable to oxygen, and placed in an Incu-shaker at 37$^{\circ}$C and 300 rpm for a night. Then, the bacteria are placed in a last growth medium (2.5 g/L of NaCl, 4 g/L of Tryptone, 4 g/L of glycerol in deionized water), and placed again in an Incu-shaker at 32$^{\circ}$C and 300 rpm for 4 hours. This medium, less rich in food, will force the bacteria to develop flagellas. Finally, using a syringe and a filter (Millex, MF millipore membrane 0.45 $\mu$m), we concentrate the bacteria and exchange the growth buffer with a motility buffer (67 mmol/L NaCl, 6.2 mmol/L K$_{2}$HPO$_{4}$, 3.8 mmol/L KH$_{2}$PO$_{4}$, and 0.9 mmol/L glucose).

\begin{figure}
	\centering
    \includegraphics[width=7.cm]{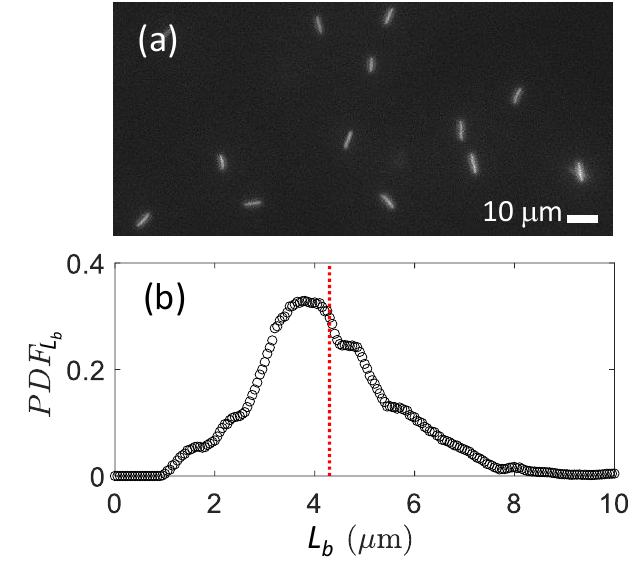}
      \caption{Bacteria length. (a) Epifluorescence microscopy image of a dilute suspension of bacteria. (b) Probability distribution function of the bacteria length, $L_{\rm b}$. The red line is the average bacteria length $\langle L_{\rm b}\rangle=4.3 \mu$m. 
}
    \label{fig:s2}
\end{figure}

\textbf{{\it E. coli} geometrical properties} --  The bacteria has a radius of $\ell_{\rm b}=$0.5$~\mu$m \cite{wu2000} and a length of $L_{\rm b}=4.3~\mu$m, Fig.~\ref{fig:s2}. 

\vspace{.1 cm}
\textbf{{\it E. coli} observation cell} -- Usually, the bacteria suspensions are enclosed between a glass slide and a cover slip, spaced by a paraffin film, in order to create a 100 $\mu m$ gap, heated to make the cover slip adherent to the glass slide. Then, the two remaining sides of the cell are sealed using NOA61 (Norland Optical Adhesives), a liquid photopolymer that cures if exposed to ultraviolet light. This sealing is compulsory to prevent any evaporation and any parasite motion. But there is a flaw: by hermetically sealing the suspension, we prevent it from receiving oxygen from outside, and as a consequence the lifetime of the bacteria is quite short. Actually, their life expectancy depends strongly on their concentration, which directly influences the global oxygen consumption. What we find out is that for small concentration, the glass slide / cover slip device is clearly sufficient, but for higher concentration, the mean velocity of the bacteria decreases from 10 $\mu$m/s to 2 $\mu$m/s in less than five minutes, which forbids us to lead experiments with a high concentrated bacteria bath. To prevent this problem we developed an agar observation cell.  Once heated in water, agar dissolves and forms, after a few minutes of cooling, a gel. As a gel, it constitutes a porous medium, and therefore lets the oxygen penetrate its structure. The porosity of the gel is related to the concentration of agar. Typically, we use a concentration of 15 g/mL. We fill a petri dish with agar dissolved in the motility buffer, and wait for the gel to solidify. Then, we extrude a cylinder of agar, and place our bacteria suspension into the well previously created. To prevent any parasite motion in the solution, we place an agar cover on the top of the well, so that the bacteria solution does not evaporate. We take care to avoid any bubble formation under the cover. 

\begin{figure}
	\centering
    \includegraphics[width=8.cm]{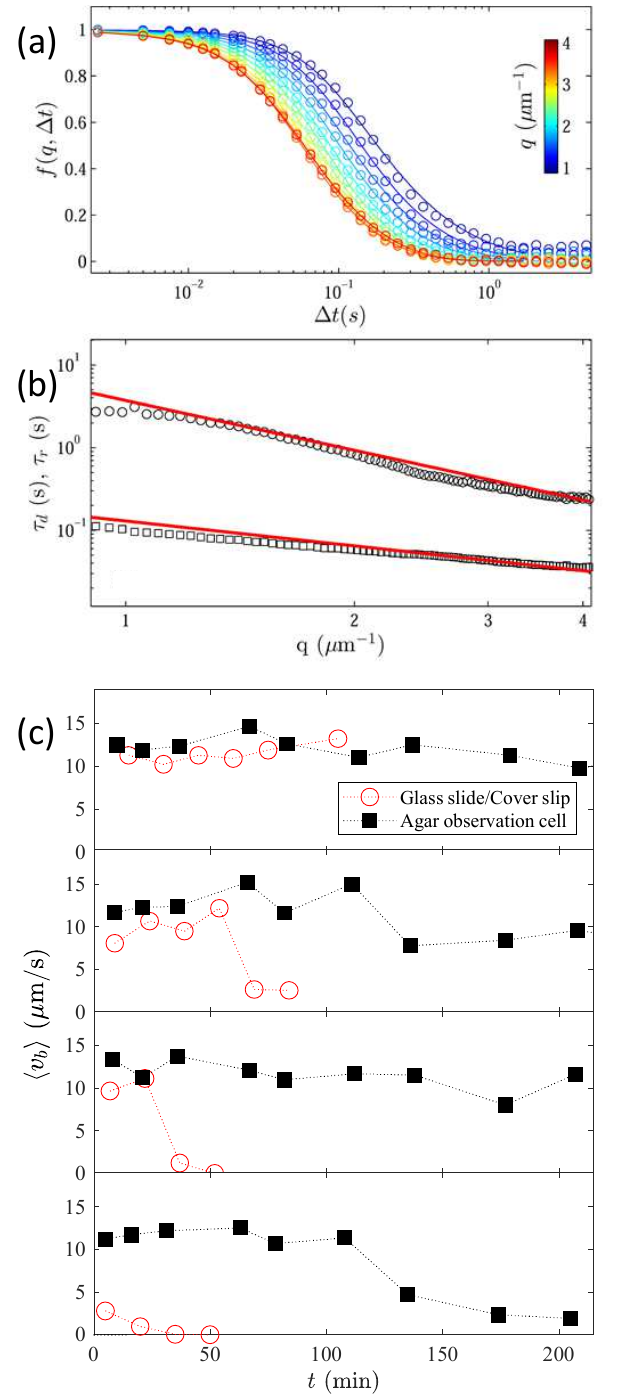}
      \caption{DDM experiment on bacteria suspensions of concentration $c$.
      (a) Auto-correlation function $f$ calculated from the DDM experiment (circles) and its fit (solid line) for various scattering number $q$. $c=1.8$ $10^{10}$ mL$^{-1}$. (b) Characteristic diffusion (circles) and ballistic (square) time extracted from the fit of $f$ in (a). The average velocity is related to the  slope of ballistic time scale, here $\langle v_{\rm b} \rangle$=11 $\mu$m/s. (c) Comparison of the average velocity of bacteria for four different concentration $c=1.8$, 4.5, 9 and 18 $10^{10}$ mL$^{-1}$ (top to bottom) between the agar observation cell and conventional glass cell.
}
    \label{fig:s1}
\end{figure}

\vspace{.1 cm}
\textbf{{\it E. coli} life expectancy} -- To measure the life expectancy of the bacteria dispersions we use differential dynamic microscopy (DDM). Indeed, when the concentration of bacteria is high, typically at least $10^{9}$ mL$^{-1}$, it is impossible to focus on single trajectories in order to deduce bacteria velocity. Therefore, we need a technique to measure a mean velocity without having to look at individual motion. This is where the DDM comes into play \cite{germain2016,martinez2012}. Using DDM, we measure, as in a dynamic light scattering, the auto-correlation function $f(q,\Delta t)$ where $q$ is the scattering wave number, Fig. \ref{fig:s1}a. Fitting $f$ with the appropriate model give access to the bacteria average velocity $\langle v_{\rm b} \rangle$, Fig. \ref{fig:s1}b.  Fig. \ref{fig:s1}c shows the time evolution of $\langle v_{\rm b} \rangle$ for the glass observation cell and the agar observation cell. If there is a drop in the velocity, it means bacteria have less energy to move and starts to die. As a consequence, we can compare the effectiveness of the two setups by comparing both the mean velocity as a function of time. At low concentrations both observation cells are equivalent, and we do not face any mortality problems. Nevertheless, as we increase the concentration, we see the limitations of the hermetic glass slide / cover slip cell. In less than 30 minutes for a concentration of $15.10^{9}$ bacteria/mL, the velocity drops to almost zero, and for the higher concentrations, the sealing of the cell with the NOA61 takes too much time to make the observation of the bacteria possible before they start dying. On the contrary, the agar cell increases substantially the bacteria lifetime, and allows us to study concentrated bacteria suspensions in a stationary regime over a period of 1h for all concentration tested in this article.

\begin{figure}
	\centering
    \includegraphics[width=8.cm]{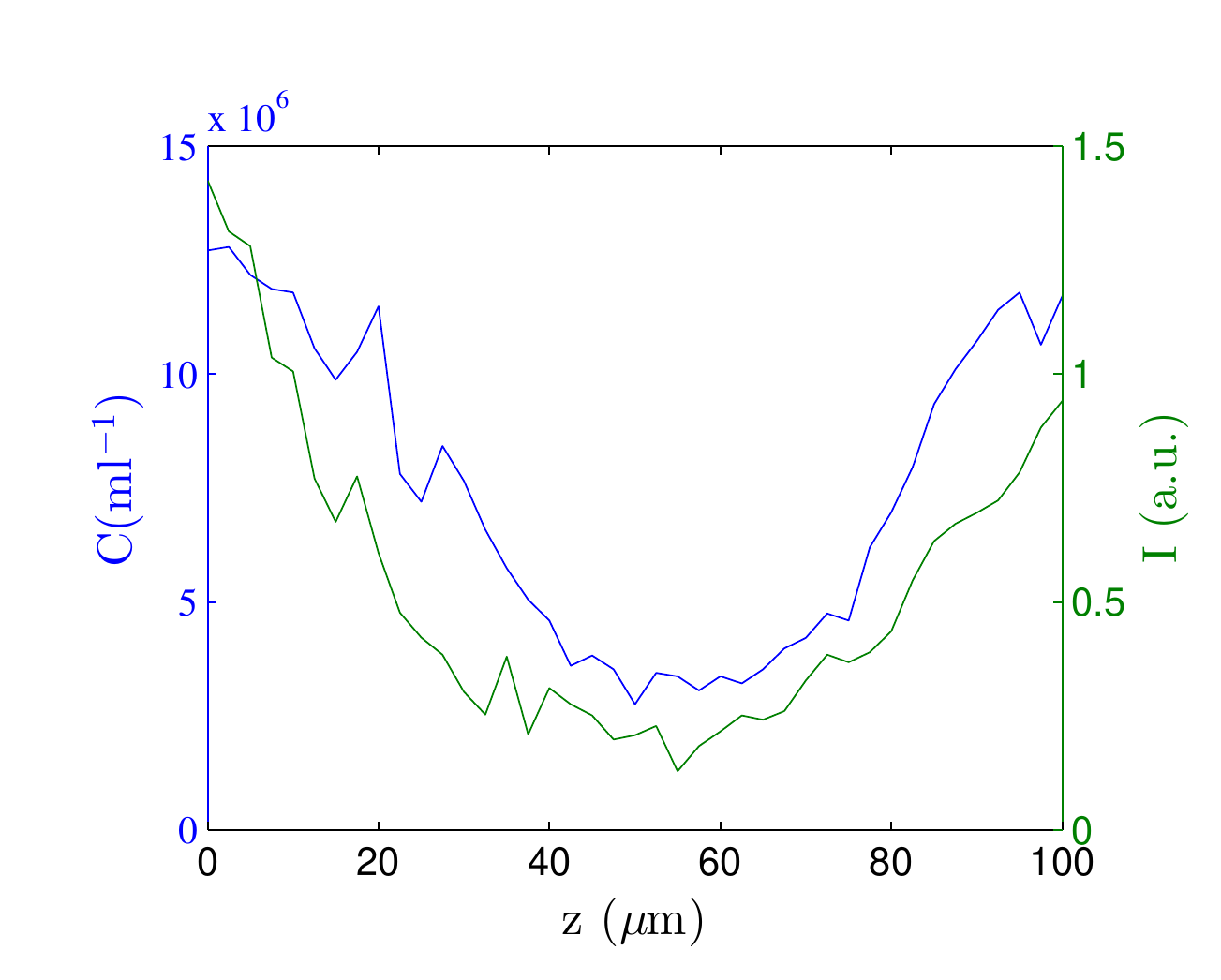}
      \caption{Bacteria concentration profile along the $z-$axis. Profile of bacteria concentration in the agar observation cell along a vertical axis ($z-$axis), measured by counting the bacteria (blue), and averaging the bacteria fluorescence intensity $I$ (green) in the same experiment.
}
    \label{fig:s3}
\end{figure}

\vspace{.1 cm}
\textbf{{\it E. coli} concentration} -- The bacteria concentration is based on the optical absorbance $OA_{600}$ at 600 nm measure with UV-visible spectrometer (ocean optics, USB4000) where $c = 1.2$ $10^{9}  OA_{600}$ \cite{schwarz2016}. $OA=\log_{10}(I_0/I)$ where $I_0$ is the solvent transmitted intensity and $I$ is the bacteria dispersion transmitted intensity. Bacteria are know to have higher density near walls. To measure the bacteria concentration profile along the $z-$direction in the agar observation cell, we use a confocal spinning disk microscope, that allows us to visualize a $z-$plan of the well  with a vertical focal depth of 8 $\mu$m, $z-$step of 2.5 $\mu$m at 0.1Hz. In Fig. \ref{fig:s3}, we use a bacteria suspension at a very low concentration, so that we are able to count individual bacteria at different heights. At the same time, we calculate the mean intensity of these very same images. As expected, we find that the fluorescent intensity is directly proportional to the number of bacteria. First of all, the profile is very similar to what we can find in literature, in \cite{berke2008} for example. We find again that bacteria are attracted to surfaces. As a consequence, we can see the concentration we have near a surface is far higher than the mean concentration we measure thanks to optical absorption. In all our experiments, we need to correct the concentration we measure by optical density by an adjustment factor to take into account this spatial inhomogeneity. As the colloids we use diffuse at the bottom surface of the agar observation cell, we measure the concentration $c$ of the bacteria in this region of interest. Experimentally, we find that the bacteria concentration is 5 times higher at the bottom of the observation cell than the average concentration measured with optical density. As a consequence, since we realize all our experiments near the surface of the agar device, we will multiply all the concentration we measure by 5.

\begin{figure}
	\centering
    \includegraphics[width=7.cm]{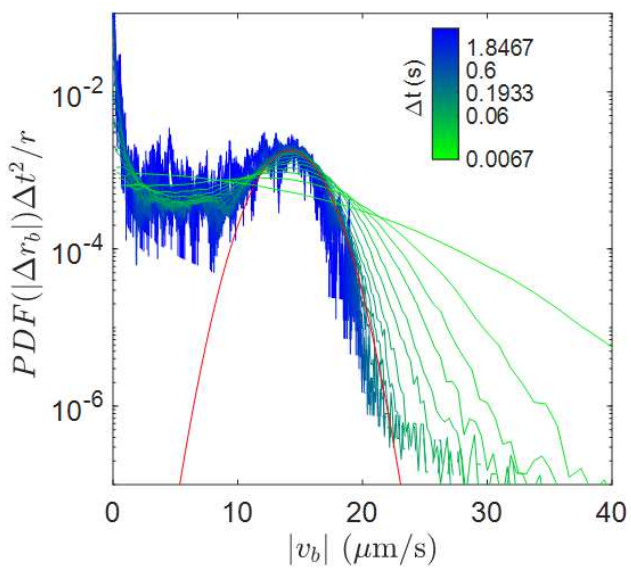}
      \caption{Bacteria velocity. Velocity distribution function as a function of $v_{\rm b}$. colors code for time. The red line is a Gaussian distribution centered on $\langle v_{\rm b}\rangle=15 \mu$m/s. The standard deviation is 4 $\mu$m/s.
}
    \label{fig:s4}
\end{figure}

\vspace{.1 cm}
\textbf{{\it E. coli} velocity} -- Single bacteria dynamics is obtained by tracking individual bacteria and identify their center of mass $r_{\rm b}(x_{\rm b},y_{\rm b})$ and their orientation defined by the angle $\theta_{\rm b}$ between the long axis of the bacteria and the $x-$axis of the laboratory. Fig. \ref{fig:s4} show the distribution function of the position increment $\mid \Delta r_{\rm b}\mid$ scaled as a function of $\mid \Delta v_{\rm b}\mid$. Measurements are averaged over 200 bacteria. For $\Delta t>0.2$ s, all curves scale on a master curve centered on Gaussian of average value $\langle v_{\rm b} \rangle=$15 $\mu$m/s with a standard deviation of 4 $\mu$m/s. This behavior is obtained on the entire range of concentrations tested in the paper. For high concentrations only a few percent of the bacteria were fluorescently labeled to allow tracking.

\vspace{.1 cm}
\textbf{Colloids distribution function} -- 
Since our problem is isotropic, the probability density function of the displacements should be of the form $H(x,y)=\bar H\left(\sqrt{x^2+y^2}\right)$.
$\bar H(r)$ is the angular average of the two-dimensional PDF $H(x,y)$; it is related to the PDF of the norm of the displacement $g(r)$, by $\bar H(r)=g(r)/(2\pi r)$.
Usually, the marginal distributions $\int H(x,y)dy$ or $\int H(x,y)dx$ are plotted. 
If $H(x,y)$ is Gaussian, the marginal distributions are also Gaussian; however, if $H(x,y)$ is not Gaussian, the marginal distributions do not represent the radial dependence of $H(x,y)$ in a straightforward way. 
For this reason, we choose here to work with $\bar H(r)$.

\begin{acknowledgments}
This work was partly supported by ANR grant StruBaDy (D.B. and T. G.). We thank Axel Buguin for sharing the bacteria strain.
\end{acknowledgments}


%

\end{document}